\documentclass[a4paper,11pt]{article}
\usepackage{pos}
\usepackage{mycommands}
\usepackage{bm}
\usepackage{bbold}
\usepackage{hyperref}

\usepackage{color}
\usepackage[compat=1.1.0]{tikz-feynman}

\title{Hubbard interaction at finite $T$ on a hexagonal lattice}

\author*[a]{Lado Razmadze}
\author[a,b]{Thomas Luu}

\affiliation[a]{Institute for Advanced Simulation (IAS-4),  Forschungszentrum J\"ulich, Germany}
\affiliation[b]{Helmholtz-Institut f\"ur Strahlen- und Kernphysik and Bethe Center for Theoretical
Physics,\\
Rheinische Friedrich-Williams-Universit\"at Bonn, Germany}

\emailAdd{l.razmadze@fz-juelich.de}
\emailAdd{t.luu@fz-juelich.de}

\abstract{The temporal finite volume induces significant effects in Monte Carlo simulations of systems in low dimensions, such as graphene, a 2-D hexagonal system known for its unique electronic properties and numerous potential applications.
In this work, we explore the behavior of fermions on a hexagonal sheet with a Hubbard-type interaction characterized by coupling $U$. This system exhibits zero or near zero-energy excitations that are highly sensitive to finite temperature effects. We compute corrections to the self-energy and the effective mass of low-energy excitations, arriving at a quantization condition that includes the temporal finite volume. These analyses are then conducted for both zero and finite temperatures. Our findings reveal that the first-order $\mathcal{O}(U)$ contributions are absent, leading to non-trivial corrections starting at $\mathcal{O}(U^2)$. We validate our calculations against exact and numerical results obtained from Hybrid Monte Carlo simulations on small lattices.
}

\FullConference{The 41st International Symposium on Lattice Field Theory (LATTICE2024)\\
 28 July - 3 August 2024\\
Liverpool, UK\\}

\begin{document}
\maketitle

\section{Introduction}
In any quantum system, we can define a characteristic temperature $T_C$, representing a scale comparable to the energy of the system's lowest eigenmode. At temperatures $T\ll T_C$, the thermal energy is insufficient to excite even the lowest eigenstates, enabling an approximation $T=0$ since temporal finite-volume effects are negligible. However, when $T\approx T_C$, the application of thermal field theory becomes essential to accurately capture the system’s behavior.

In the context of lattice QCD, this characteristic temperature corresponds to the pion mass $T\approx m_\pi\approx150\text{MeV}$, a sufficiently high threshold where thermal effects significantly impact the study of nuclear matter and quark-gluon plasma \cite{bellac, kapusta}. In ``cold'' lattice QCD calculations, where the temporal extent is $\approx 30-40\ \text{MeV}\ll m_\pi$, the finite temporal effects are justifiably ignored.  Conversely, in physical low-dimensional lattice structures—such as graphene sheets, graphene nanoribbons, and topological insulators—the lowest energy eigenmodes are much lower than the temporal extent, often just a few eV or even zero~\cite{CastroNeto:2007fxn}, e.g. the momentum $K$-points in graphene. In these systems thermal effects must be included from the outset.

While temperature-dependent properties of small lattices can be investigated using computational approaches like Hamilonian Monte Carlo (HMC), these methods remain computationally demanding even for modest lattice sizes. In this work, we approach the problem analytically, solving the self-energy $\Sigma$ of the system perturbatively in the Hubbard coupling $U$ to arrive at explicit temperature dependencies. For small lattices, exact solutions are also possible, enabling a direct comparison with perturbative results. Throughout this paper, we use the inverse temperature notation, $\beta=1/T$, to express temperature-dependent quantities.

\subsection{System}
\begin{figure}[h]
    \centering
    \includegraphics[width=.55\textwidth, trim={0 13cm 0 0}]{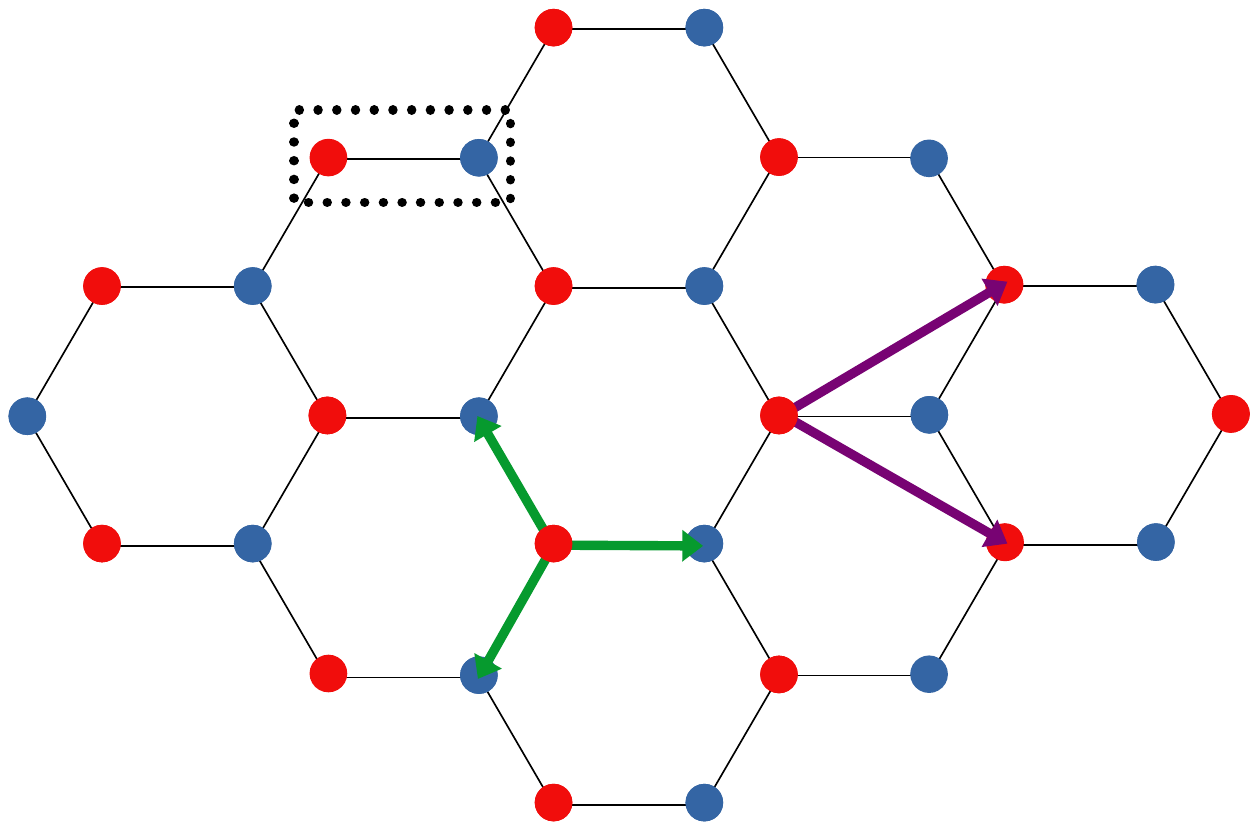}
    \caption{$4\times4$ Grahene sheet. Graphene forms a hexagonal bipartite lattice with two sites {\color{red}A} and {\color{blue}B} in each unit cell (dotted rectangle), located at $\xi_A = (0,0)$ and $\xi_B = (1,0
    )$ respectively. The lattice translation vectors, shown in purple, are $a_{1,2} = (3/2, \pm\sqrt{3}/2)$, and the nearest-neighbor vectors are marked by green arrows.}
    \label{fig:graphene}
\end{figure}
\begin{figure}[h]
    \centering
    \includegraphics[width=\textwidth, trim={0 5mm 0 2mm}]{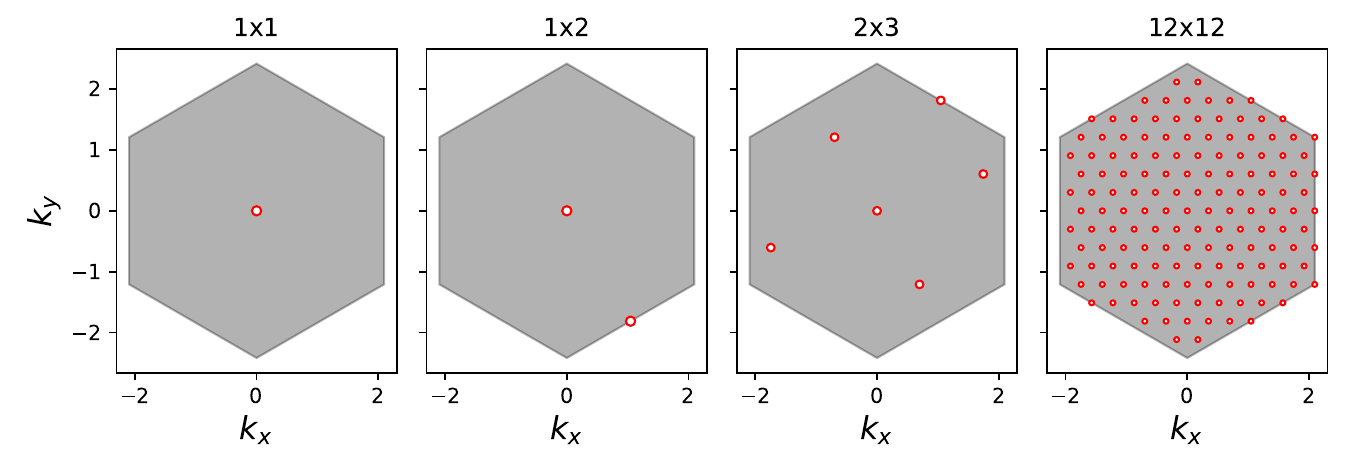}
    \caption{Brillouin zones for various lattice sizes in graphene. As lattice size increases, momentum points increasingly populate the hexagonal Brillouin zone. Of particular importance are the momentum points at the center ($\Gamma$-point), at the midpoint of the edges ($M$-point), and at the vertices ($K$-points). In the thermodynamic limit when $L_x,L_y\to\infty$, the BZ forms a hexagon.}
    \label{fig:brillouin_zone}
\end{figure}

We investigate this system using a tight-binding Hamiltonian with an on-site Hubbard interaction term at half-filling~\cite{Ostmeyer:2020uov}, given by: 
\begin{equation}	
    H = 
    -\sum_{\< x,y \>s} (c_{xs}\+ c^{}_{ys} + c_{ys}\+ c^{}_{xs}) 
    -\frac{U}{2}\sum_x(n_{x\up} - n_{x\dn})^2
\end{equation}
where $\langle x,y \rangle$ indicates a summation over nearest-neighbor sites, $c\+/c$ denotes the fermionic creation/annihilation operator, and $n=c\+c$ represents the corresponding number operator.
This Hamiltonian is applied to a graphene sheet of size $L_x\times L_y$ (see Fig.~\ref{fig:graphene}), where $L_x$ and $L_y$ refer to the width and length of the sheet in unit cells. We will refer to the $1\times1$ sheet as the 2-site and the $1\times2$ as the 4-site lattice. 
The graphene structure also has a well-defined Brillouin zone (BZ) comprising of $\L=L_xL_y$ momentum points.  Examples are given in Fig.~\ref{fig:brillouin_zone}.

To diagonalize the Hamiltonian, we apply a Fourier transform on each sublattice. Letting $x$ enumerate unit cells and $\lam$ denote the sublattice, we define:
\begin{equation}
    c^{}_{x\lam s} = \inv{\sqrt{\Lambda}}\sum_{k} e^{ik(x+\xi_\lam)} c^{}_{k\lam s}\ .
\end{equation}
The Fourier transform yields a 2x2 matrix in sublattice space, which is diagonalized to arrive at
\begin{equation}\label{eq:H0}
    H_0 = \sum_{k\rho s}
    \E_k^\rho \phi\+_{k\rho s} \phi^{}_{k\rho s},
    \quad
    \E_k^\rho = \rho \E_k = \rho |f(k)|
\end{equation}
where 
\begin{equation}
    f(k) = e^{ik_x} + 2e^{-ik_x/2}\cos(\sqrt{3}k_y/2)
    \quad\text{and}\quad
    \rho=\pm1\,.
\end{equation}
To clarify the physical meaning of $\rho$, we introduce the Fermi energy $\E_F$, defined such that all states with $\E^\rho_k \leq \E_F$ are filled \cite{fw}. States above $\E_F$ are called \emph{particles} and states below $\E_F$ are called \emph{holes}. In our case $\E_F=0$ which means $\rho=\pm1$ states represent particles/holes.

In this representation the interaction part of the Hamiltonian can be split into a quadratic part, $H_1$, which we will refer to as the \emph{mass term} and a quartic part, $H_2$, which we will call the \emph{interaction vertex}. The mass term $H_1$ is then
\begin{equation}\label{eq:H1}
    H_1 = - \frac{U}{2}\sum_{k\rho s}
    \phi\+_{k\rho s} \phi^{}_{k\rho s}
\end{equation}
For $H_2$ we introduce the multi-index notation $\bm{k} = (k, \rho)$ and define the quartic interaction vertex
\begin{equation}
    V_{\bm{k'}\bm{l'}\bm{k}\bm{l}}
    =
    \frac{U}{4\Lambda}
    \d_{k'+l',k+l}
    \left(1 +
    \rho'\s'\rho\s
    e^{-i(\th_{k} - \th_{k'} + \th_{l} - \th_{l'})}
    e^{-i(k'+l'-k-l)\xi_b}
    \right),
    \quad
    e^{i\th_k} = \frac{f(k)}{|f(k)|}
\end{equation}
where $\d_{k'+l',k+l}$ is defined modulo the BZ. Then the final Hamiltonian takes the following form
\begin{equation}\label{eq:H2}
    H = \sum_{\bm{k} s}
    \left(\E_{\bm{k}} - \frac{U}{2}\right) 
    \phi\+_{\bm{k}s} \phi_{\bm{k}s}
    +
    \sum_{\bm{k'}\bm{l'}\bm{k}\bm{l}}
    V_{\bm{k'}\bm{l'}\bm{k}\bm{l}}
    \phi\+_{\bm{k'}\up}
    \phi\+_{\bm{l'}\dn}
    \phi^{}_{\bm{k}\dn}
    \phi^{}_{\bm{l}\up}
\end{equation}
\section{Thermal Field Theory}
We operate within the interaction picture in \emph{imaginary time} $\tau\equiv i t$, and we perturb about the free, non-interacting ($U=0$) system.
We designate the $U$-independent part of the Hamiltonian as the free Hamiltonian, $H_0$, and denote the remainder as the interaction term, $H_I$. Our free propagator is
\begin{equation}
    G^0_{\bm{k} s}(\tau) = -\left\< T_\tau [\phi^{}_{\bm{k} s}(\tau) \phi\+_{\bm{k} s}] \right\>_0
\end{equation}
where $\left\< T_\tau [\cdots] \right\>_0$ stands for \emph{thermal average} and $T_\tau$ is a \emph{time ordering} operator. The explicit time-dependent form of this quantity is:
\begin{equation}
    G^0_{\bm{k}}(\tau) 
    =
    e^{-\tau \E_{\bm{k}}}
    \begin{cases}
        -(1-n_{\bm{k}}) & \tau > 0\\
        n_{\bm{k}} & \tau \leq 0
    \end{cases}
    \qquad\text{with}\qquad
    n_{\bm{k}} = \inv{e^{\beta \E_{\bm{k}}}+1}
\end{equation}
where $n_{\bm{k}}$ is the fermion number. Since $G^0$ is spin independent we drop the $s$. This propagator is periodic in imaginary time, allowing us to express it in terms of a Fourier transform in $\tau$,
\begin{equation}
    G^0_{\bm{k}}(i\omega) 
    =
    \int_{0}^{\beta} d\tau G^0_{\bm{k}}(\tau) e^{i\omega\tau}
    =
    \inv{i\omega-\E_{\bm{k}}},
    \qquad 
    \omega=\frac{\pi}{\beta}(2m+1)\quad m\in\mathbb{Z}\ .
\end{equation}
The inverse transform is provided by the \emph{Matsubara sum}
\begin{equation}
    G^0_{\bm{k}}(\tau) 
    =
    \inv{\beta} \sum_{\omega} \frac{e^{-i\omega\tau}}{i\omega-\E_{\bm{k}}}\ .
\end{equation}
To account for interactions, we introduce a quantity analogous to the $S$-matrix in ordinary QFT \cite{abgd}, allowing us to express the interacting propagator which we'll simply refer to as a \emph{correlator}. 
\begin{equation}
    C_{\bm{k}}(\tau) = 
    -\frac{\left\<T_\tau[S(\beta)\phi_{\bm{k} s}(\tau)\phi\+_{\bm{k} s}]\right\>_0}
    {\left\<S(\beta)\right\>_0},
    \qquad
    S(\tau)
    =
    T_\tau \exp\left\{-\int_0^\tau H_I(\tau') d\tau' \right\}
\end{equation}
According to Wick’s theorem, thermal average of time-ordered product of the fermionic operators can be expressed as the sum of all possible \emph{Wick contractions}, defined as 
\begin{equation}
    \phi^\bullet_{\bm{k}s}(\tau_1)\phi^{\dagger\bullet}_{\bm{l}s'}(\tau_2) = -\d^{}_{\bm{k}\bm{l}}\d^{}_{ss'} G^0_{\bm{k}}(\tau_1 - \tau_2)\,.
\end{equation}
Rather than computing each contraction explicitly, we employ a diagrammatic approach, using Feynman diagrams to systematically capture higher-order contributions. In this case, we have one propagator and two interaction vertices from $H_1$ and $H_2$, respectively. The complete set of Feynman rules for our calculations is as follows:
\begin{equation}
	\begin{matrix}
        \begin{matrix}
            \begin{tikzpicture}
                \begin{feynman}
                    \vertex (i);
                    \vertex (o) at ($(i)+(2,0)$);
                    \diagram*{
                        (i) -- [fermion] (o);
                    };
                \end{feynman}
            \end{tikzpicture}
            &\sim& \inv{i\omega - \E_{\bm{k}}} \equiv G^0_{\bm{k}}(i\omega) \\ \\
            \begin{tikzpicture}
                \begin{feynman}
                    \vertex (i);
                    \vertex [square dot] (m) at ($(i)+(1,0)$) {};
                    \vertex (o) at ($(i)+(2,0)$);
                    \diagram*{
                        (i) -- (m) -- (o);
                    };
                \end{feynman}
            \end{tikzpicture}
            &\sim&
            -\frac{U}{2}
        \end{matrix} &\qquad&
        \begin{tikzpicture}[baseline=(o.base)]
            \begin{feynman}
                \vertex [dot] (o) {};
                \vertex (i1) at ($(o)+(-1,-.707)$) {$\bm{k}\up$};
                \vertex (i2) at ($(o)+(-1,.707)$) {$\bm{l}\dn$};
                \vertex (i3) at ($(o)+(1,-.707)$) {$\bm{k'}\up$};
                \vertex (i4) at ($(o)+(1,.707)$) {$\bm{l'}\dn$};
                \diagram*{
                    (i1) -- [fermion] (o) -- [fermion] (i3);
                    (i2) -- [fermion] (o) -- [fermion] (i4);
                };
            \end{feynman}
        \end{tikzpicture}
        &\sim& V^{}_{\bm{k'}\bm{l'}\bm{l}\bm{k}}
	\end{matrix}
\end{equation}
In this way, we only have to keep track of topologically distinct Feynman diagrams instead of all the possible Wick contractions.
\section{Solution}
In order to obtain a non-perturbative correction to the propagator we utilize the \emph{self-energy} $\Sigma_{\bm{k}}(i\omega)$, which is the sum of all 1 particle irreducible (1-PI) diagrams.
$\Sigma$ is a 2x2 matrix in the sub-lattice basis and in general not diagonal. The fully dressed correlator (double line) can then be expressed using Dyson's equation
\begin{align*}
	\begin{tikzpicture}[baseline=(i.base)]
		\begin{feynman}
			\vertex (i); 
			\vertex (o) at ($(i)+(1,0)$);
			\diagram*{
				(i) -- [double] (o);
			};
		\end{feynman}
	\end{tikzpicture}
	\quad&=\quad
	\begin{tikzpicture}[baseline=(i.base)]
		\begin{feynman}
			\vertex (i); 
			\vertex (o) at ($(i)+(1,0)$);
			\diagram*{
				(i) -- [fermion] (o);
			};
		\end{feynman}
	\end{tikzpicture}
	\quad+\quad
	\begin{tikzpicture}[baseline=(i.base)]
		\begin{feynman}
			\vertex (i); 
			\vertex [blob] (p) at ($(i)+(1,0)$) {$\Sigma$};
			\vertex (o) at ($(i)+(2,0)$);
			\diagram*{
				(i) -- [fermion] (p) -- [double] (o);
			};
		\end{feynman}
	\end{tikzpicture}\ .
\end{align*}
Poles of the correlator provide the spectrum arising from self-interactions, leading to the \emph{quantization condition} (QC).
\begin{equation}\label{eq:quantization-condition}
	\det\left(G^0_{\bm{k}}(i\omega)^{-1} - \Sigma^{}_{\bm{k}}(i\omega)\right)\bigg|_{i\omega=E_{\bm k}} = 0
\end{equation}
where $E_{\bm{k}}$ is the interacting energy. Note that all temperature dependence (ie $\beta$-dependence) originates in $\Sigma$. To get the time-dependent correlator $C_{\bm{k}}(\tau)$ we must perform a Matsubara sum. If $\Sigma$ is diagonal we use standard complex integration techniques to rewrite the sum over frequencies $\omega$ as a contour integral which can be solved as a sum over the residues at the poles of the correlator \cite{fw}. 
\begin{equation}\label{eq:second-order-correlator}
   C_{\bm k}(\tau)= \inv{\beta}\sum_\omega\frac{e^{-i\omega \tau}}{i\omega - \E_{\bm{k}} - \Sigma_{\bm{k}}(i\omega)} 
    =
    \sum_{z^*}
    \frac{e^{-z^*\tau}}{e^{-z^*\beta}+1}
    \text{Res}\left(\inv{z - \E_{\bm{k}} - \Sigma_{\bm{k}}(z)}, z=z^*\right)
\end{equation}
 Even if $\Sigma$ is only approximately diagonal we can assume the off-diagonals to be zero with negligible error. We have also observed that for $\Gamma$, $M$ and $K$ points self-energy is always diagonal.

We calculate $\Sigma$ to the leading non-trival order in $U$. At $\mathcal{O}(U)$ there are only two contributions, one from the mass term and the other from the interaction vertex. These contributions are constant and opposite in sign, canceling each other exactly. Diagramatically the cancellation is given as 
\begin{equation}\label{eq:U1cancel}
    \begin{tikzpicture}
        \begin{feynman}
            \vertex (i);
            \vertex [dot] (x1) at($(i)+(1,0)$) {};
            \vertex (x2) at($(i)+(1,.8)$);
            \vertex (o) at ($(i)+(2,0)$);
            \diagram*{
                (i) -- [fermion] (x1) -- [fermion] (o);
                (x1) -- [fermion, half left] (x2) -- [fermion, half left] (x1);
            };
        \end{feynman}
    \end{tikzpicture}
    +
    \begin{tikzpicture}
        \begin{feynman}
            \vertex (i);
            \vertex [square dot] (m) at ($(i)+(1,0)$) {};
            \vertex (o) at ($(i)+(2,0)$);
            \diagram*{
                (i) -- (m) -- (o);
            };
        \end{feynman}
    \end{tikzpicture}
    = 0
\end{equation}
This implies that the leading order contribution must be at least $\sim\mathcal{O}(U^2)$. At this order we can use \eqref{eq:U1cancel} and assume all diagrams containing ``tadpoles" and mass terms to cancel and we are left with a single ``sunset diagram":
\begin{equation}\label{eq:second-order-sigma}
\begin{tikzpicture}[baseline=(o.base)]
    \begin{feynman}
        \vertex (i) {$p\pi$};
        \vertex [dot] (x1) at($(i)+(1.2,0)$) {};
        \vertex [dot] (x2) at($(i)+(2,0)$) {};
        \vertex (o) at ($(i)+(3.2,0)$) {$p\pi$};
        \diagram*{
            (i) -- [fermion] (x1) -- [fermion] (x2) -- [fermion] (o);
            (x1) -- [fermion, half left] (x2) -- [fermion, half left] (x1);
        };
    \end{feynman}
\end{tikzpicture}        
=
\sum_{\bm{l'}\bm{k}\bm{l}}
|V^{}_{\bm{p}\bm{l'}\bm{k}\bm{l}}|^2
\frac{n_{-\bm{l'}}n_{\bm{k}} + (n_{\bm{l'}} - n_{\bm{k}})n_{\bm{-l}}}
{i\omega - (\E_{\bm{k}} - \E_{\bm{l'}} + \E_{\bm{l}})}
\end{equation}
where we used shorthand $-\bm{k}=(k,-\rho)$, and also defined $\bm{p}=(p,\pi)$ for external lines.

For the 2-site lattice there is only one momentum point $\Gamma$, therefore $\E_k^\rho = \rho\E_\Gamma$. Here $|V_{\bm{p}\bm{l'}\bm{k}\bm{l}}|^2$ simplifies to $\frac{U^2}{4}\d_{\pi\s',\rho\s}$. At zero temperature the QC becomes a quadratic equation:
\begin{equation}
    (E^\pi_\Gamma - \pi\E_\Gamma)(E^\pi_\Gamma + 3\pi\E_\Gamma) -  \frac{U^2}{4} = 0
\end{equation}
With the additional condition that $E^\pi_\Gamma\big|_{U=0} = \pi\E_\Gamma$ we can analytically solve for the energy
\begin{equation}\label{eq:2-site-T0-energy}
    E^\pi_\Gamma = \pi\E_\Gamma \left(\sqrt{\frac{U^2}{4\E^2_\Gamma}+4} - 1\right)
\end{equation}
Even at $T\neq0$ the QC for 2-sites is a cubic polynomial, making it anaytically solvable. However, exact expressions tend to be cumbersome and do not offer substantial insight. For larger lattices, solving the QC analytically becomes infeasible so we can use numerical methods to find the roots of \eqref{eq:quantization-condition}. Due to the form of \eqref{eq:second-order-sigma}, QC is always a rational function of $E_{\bm{k}}$, meaning we can also find residues and perform the summation in \eqref{eq:second-order-correlator}. For larger lattices, however, accumulation of numerical errors precludes us from performing the residue sum to a sufficient degree of accuracy, forcing us to fall back on HMC. In principle, one could use arbitrary precision methods, albeit with a significant performance hit.

\section{Results}
\subsection{$T=0$}
For the 2-site and 4-site systems, we compare perturbative results with exact solutions to assess the validity of the leading-order (LO) approximation. First, in the zero-temperature limit ($\beta\to\infty$), we examine the dependence of energy on $U$ alone. In the 2-site case perturbative expression \eqref{eq:2-site-T0-energy} is equal to the exact solution given in \cite{hubbard}. For the 2-site problem, higher-order self-energy contributions vanish at $T=0$, making it exactly solvable in our formalism.

\begin{figure}[h]
    \centering
    \includegraphics[width=.95\textwidth, trim={0 1.5cm 0 0}]{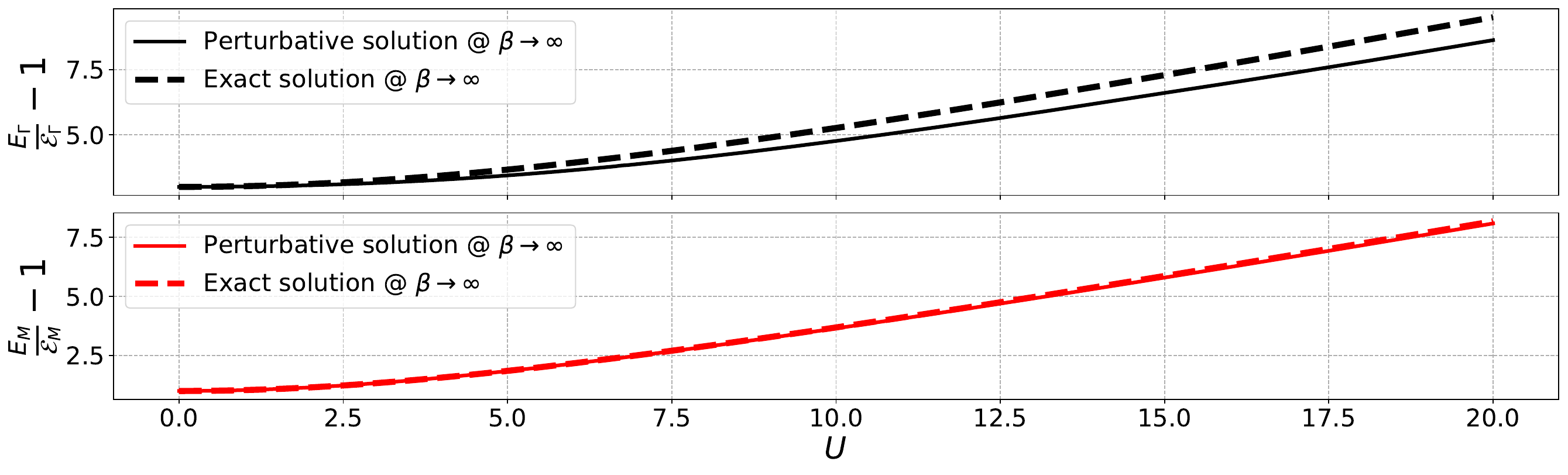}
    \caption{Energy dependence on $U$ for the 4-site system at $\beta\to\infty$. Plots are shown for $\Gamma$ and $M$ points, demonstrating good agreement between perturbative and exact solutions even for large values of $U$.}
    \label{fig:energy 4-site}
\end{figure}

For the 4-site system, which includes two momentum points in the Brillouin Zone (BZ) -- $\Gamma$ and $M$ -- the discrepancy between perturbative and exact results is more pronounced. Nevertheless, the agreement remains excellent for both momentum points, even up to $U=20$, which is well into the non-perturbative regime, see Fig. \ref{fig:energy 4-site}.

\subsection{$T\neq0$}
To examine finite-temperature behavior, we calculate the time-dependent correlators using equation \ref{eq:second-order-correlator}. Notably, the particle and hole propagators are mirror images around $\tau=\beta/2$. For the 2-site system, the perturbative and exact calculations are in close agreement, although deviations become noticeable at very high values of $U$ due to higher-order contributions at finite temperatures.

\begin{figure}[h]
    \centering
    \includegraphics[width=\textwidth, trim={0 5mm 0 0}]{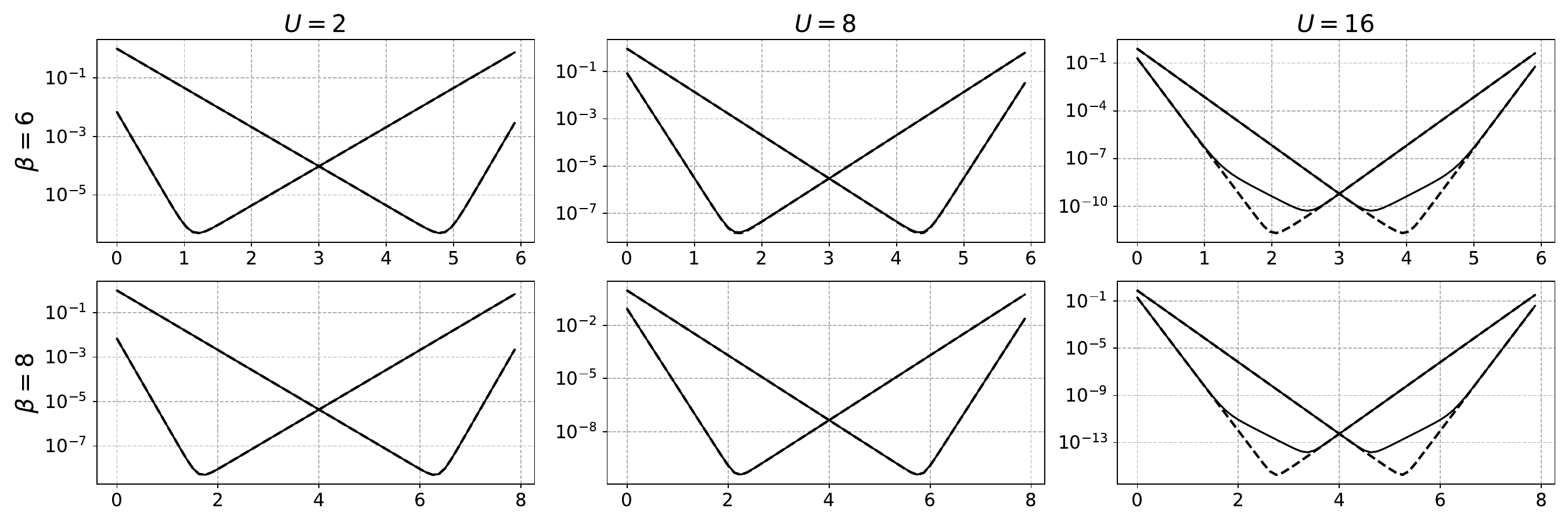}
    \caption{$C_{\bm{k}}(\tau)$ vs. $\tau$ for the 2-site system with $U=2,8,16$ and $\beta=8,10$. Deviations between perturbative (dashed) and exact (solid) results only become apparent at very large $U$. Even at $U=8$, a strongly non-perturbative value, the results show close agreement.}
    \label{fig:corelators 2-site}
\end{figure}

For the 4-site system, we analyze correlators at the $\Gamma$ and $M$ points. The deviation between perturbative and exact results is more pronounced here than in the 2-site case. However, the agreement is still good, extending beyond typical perturbative limits of $U$.

\begin{figure}[h]
    \centering
    \includegraphics[width=\textwidth, trim={0 5mm 0 0}]{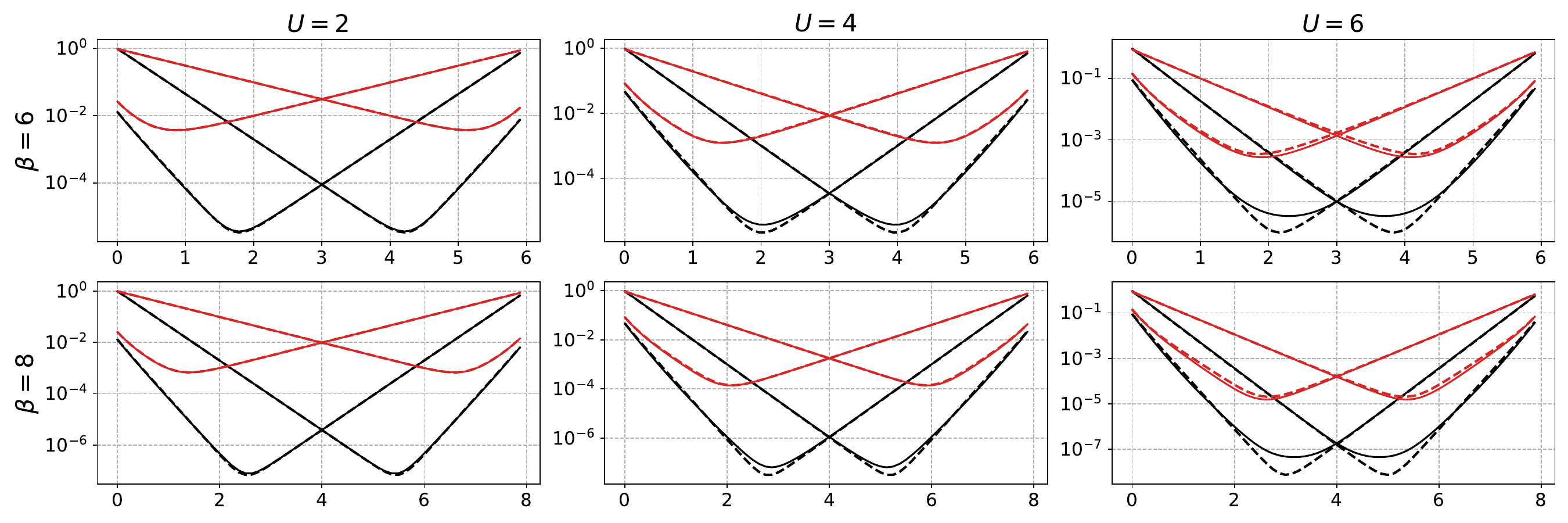}
    \caption{$C_{\bm{k}}(\tau)$ vs. $\tau$ for the 4-site system at $\Gamma$ (black) and $M$ (red) points, with $U=2,4,6$ and $\beta=6,8$. Deviations between perturbative (dashed) and exact (solid) results are more apparent than in the 2-site case but still show strong agreement for high values of $U$.}
    \label{fig:corelators 4-site}
\end{figure}

\begin{figure}[h]
    \centering
    \includegraphics[width=\textwidth, trim={0 5mm 0 0}]{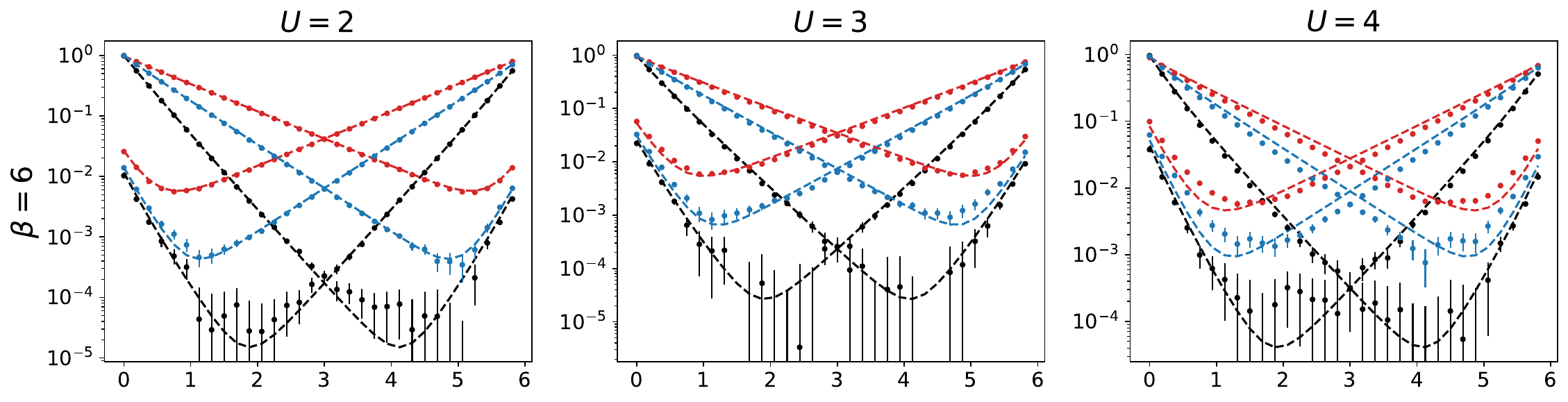}
    \caption{$C_{\bm{k}}(\tau)$ vs. $\tau$ for $2\times3$ graphene at $U=2$, $3$, and $4$, all with $\beta=6$. These are $\Gamma$ (black), $M$ (red) and $A$ (blue) points. Dashed lines are perturbative correlators and the dots are from an HMC simulaton. Agreement is very good. Even for $A$ which is not diagonal in particle-hole basis. However, off-diagonal terms are small enough to treat it as approximately diagonal. To observe the deviation one has to go up to $U=4$.}
    \label{fig:correlators 2x3}
\end{figure}

For lattices larger than 4-sites, exact solutions become computationally prohibitive, necessitating the use of Hybrid Monte Carlo (HMC) simulations~\cite{Ostmeyer:2021efs,Smith:2014tha,Ulybyshev:2013swa}. In Fig. \ref{fig:correlators 2x3}, we present the correlators for a $2\times3$ graphene. Although the Brillouin Zone (BZ) contains 6 momentum points, only 3 unique correlators emerge, labeled as $\Gamma$, $M$, and $A$. The latter correlator provides an example where the self-energy matrix $\Sigma$ is not strictly diagonal.
Despite this, we can approximate a solution to the quantization condition. Our calculations reveal that the off-diagonal terms are an order of magnitude smaller than the diagonal terms. If we denote this difference by $\epsilon$, by setting these off-diagonal terms to zero, we incur an error on the order of $\epsilon^2$, which remains negligible in this context.

As in the 2- and 4-site examples, we find very good agreement between our perturbative and HMC calculation for couplings up to $U=3$.  However, at $U=4$ we see definitive discrepancies in the correlators.
\section{Summary and Outlook}
This proceeding presents perturbative calculations of the self-energy $\Sigma$ in thermal field theory applied to the Hubbard model on a graphene lattice, computed up to the leading non-trivial order, $\mathcal{O}(U^2)$. We have calculated the zero-temperature energy shift, achieving perfect agreement with exact results for the 2-site system and strong agreement for the 4-site system, even for non-perturbative values as high as $U=20$. We further analyzed the time evolution of correlators for both 2-site and 4-site models, observing remarkable alignment between perturbative and exact results despite the large $U$ values. Additionally, we investigated correlators for $2\times3$ graphene sheet and compared our calculations to HMC simulated data for up to $U=3$ demonstrating good consistency. Only at larger $U$ do we see a discrepancy between perturbative and HMC results. In the future we will use our formalism here to deduce the finite-temperature dependence of eigen-energies obtained from HMC simulations within a finite temporal volume. 

\section*{Acknowledgements}
We thank Evan Berkowitz for insightful discussions and Petar Sinilkov for providing routines to obtain the exact solutions.
We gratefully acknowledge the computing time granted by the JARA Vergabegremium and provided on the JARA Partition part of the supercomputer JURECA at Forschungszentrum J\" ulich \cite{jara}.

\end{document}